\pdfoutput=1

\documentclass[11pt]{article}

\usepackage{acl}
\usepackage{graphicx}
\usepackage{times}
\usepackage{latexsym}

\usepackage{multirow}

\usepackage[T1]{fontenc}

\usepackage[utf8]{inputenc}

\usepackage{microtype}

\usepackage{inconsolata}
\usepackage{booktabs}
\usepackage{amsmath}
\usepackage{soul}

\title{SciMMIR: Benchmarking Scientific Multi-modal Information Retrieval
}

\newcommand*\samethanks[1][\value{footnote}]{\footnotemark[#1]}
\author{
\small
    Siwei Wu\textsuperscript{\includegraphics[scale=0.03]{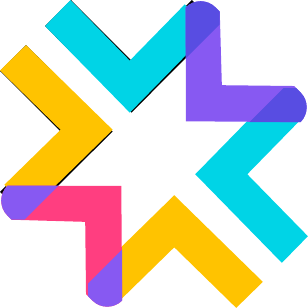},1}\quad 
    Yizhi Li\textsuperscript{\includegraphics[scale=0.03]{figures/map-logo-c.pdf},1}\quad 
    Kang Zhu\textsuperscript{\includegraphics[scale=0.03]{figures/map-logo-c.pdf}}\quad 
    Ge Zhang\textsuperscript{\includegraphics[scale=0.03]{figures/map-logo-c.pdf},2}\quad 
    \textbf{Yiming Liang}\textsuperscript{3}
\\
\small
    \textbf{Kaijing Ma}\textsuperscript{4}\quad
    \textbf{Chenghao Xiao}\textsuperscript{\includegraphics[scale=0.03]{figures/map-logo-c.pdf}, 5}\quad
    \textbf{Haoran Zhang}\textsuperscript{\includegraphics[scale=0.03]{figures/map-logo-c.pdf}, 4}\quad
    \textbf{Bohao Yang}\textsuperscript{1}
\\
\small
    \textbf{Wenhu Chen}\textsuperscript{\includegraphics[scale=0.03]{figures/map-logo-c.pdf},2}\quad
    \textbf{Wenhao Huang}\textsuperscript{\includegraphics[scale=0.03]{figures/map-logo-c.pdf},3}\quad 
    \textbf{Noura Al Moubayed}\textsuperscript{5}\quad
    \textbf{Jie Fu}\textsuperscript{\includegraphics[scale=0.03]{figures/map-logo-c.pdf},4}\thanks{Corresponding authors.}\quad
    \textbf{Chenghua Lin}\textsuperscript{\includegraphics[scale=0.03]{figures/map-logo-c.pdf},1}\samethanks[1]
\\
\small
    \textsuperscript{\includegraphics[scale=0.03]{figures/map-logo-c.pdf}}Multimodal Art Projection Research Community\quad
    \textsuperscript{1}University of Manchester\quad
    \textsuperscript{2}University of Waterloo
\\
\small
    \textsuperscript{3}01.ai\quad
    \textsuperscript{4}Hong Kong University of Science and Technology\quad
    \textsuperscript{5}Durham University
}

\begin{document}

\newcommand{\fwdtask}{{\texttt{txt}$\rightarrow$\texttt{img}}}
\newcommand{\baktask}{{\texttt{img}$\rightarrow$\texttt{txt}}}

\maketitle

\begin{abstract}

Multi-modal information retrieval (MMIR) is a rapidly evolving field where significant progress has been made through advanced representation learning and cross-modality alignment research, particularly in image-text pairs.
However, current benchmarks for evaluating MMIR performance on image-text pairs overlook the scientific domain, which has characteristics that are distinct from generic data, as the captions of scientific charts and tables usually describe experimental results or scientific principles, rather than human activity or scenery.
To bridge this gap, we develop a \textbf{sci}entific domain-specific \textbf{MMIR} benchmark (\textbf{SciMMIR}) by leveraging corpora of open-access research papers to extract data relevant to the scientific domain. 
This benchmark comprises \textbf{530K} meticulously curated image-text pairs extracted from figures and tables with detailed captions from scientific documents.
We further annotate the image-text pairs with a two-level subset-subcategory hierarchy to facilitate a more comprehensive evaluation of baseline retrieval systems. 
We conduct zero-shot and fine-tuned evaluations on prominent multi-modal image-captioning and visual language models, such as CLIP, BLIP, and BLIP-2.
Additionally, we perform optical character recognition (OCR) on the images and exploit this text to improve the capability of VLMs on the SciMMIR task.
Our findings offer useful insights for MMIR in the scientific domain, including the influence of pre-training and fine-tuning settings, the effects of different visual and textual encoders, and the impact of OCR information. 
All our data and code are made publicly available.\footnote{\url{https://github.com/Wusiwei0410/SciMMIR}}

\end{abstract}

\section{Introduction}

Information retrieval (IR) systems are expected to provide a relevant piece of information from a vast, yet organised, collection of data, according to given user queries. 
With the advancement of representation learning~\cite{bengio2013representation}, the methodological paradigm of IR systems has evolved from using lexical matching to retrieve textual data~\cite{luhn1957statistical, jones2000probabilistic, robertson2009probabilistic} to a mixture of similarity matching approaches in a learned representation space, consequently supporting additional modalities such as images and audio, alongside text~\cite{karpukhin2020dense,chen2020uniter,koepke2022audioretrieval}.

In scientific domains, offering users a fine-grained multi-modal retrieval service presents considerable practical significance.
Although previous studies have evaluated the image-text retrieval task across a range of general topics on large-scale datasets such as Wikipedia~\cite{young2014image, lin2014mscoco, srinivasan2021wit,goldsack2023domain,luo2023ReMuQ}, there is a notable research gap in comprehensively assessing MMIR models within scientific domains, specifically.
Integrating both in-domain and out-of-domain data in the pre-training phase significantly boosts the performance of visual language models (VLMs) on downstream tasks. However, the training of most VLMs has focused exclusively on common generic topics concerning the mundane events of daily life \cite{luo2023end}, such as images depicting scenery and human activities. As a result, this pre-training overlooks data pertinent to scientific domains such as elements related to model architectures, illustrations of scientific principles, and the results of experiments. 

Due to the substantial differences in the data distribution characteristics between generic data and scientific data, many VLMs may not have an adequate ability to perform MMIR for scientific domains.
Additionally, existing table-related works, such as table generation tasks, have mainly focused on textual representations of tables, while overlooking image-based representations of tabular data.
This presents problems for human-computer interaction, as users may desire to input information in the form of screenshots and expect an interactive system to present results in a graphical format.

\begin{figure*}[!tb]
    \centering
    \includegraphics[width=14cm]{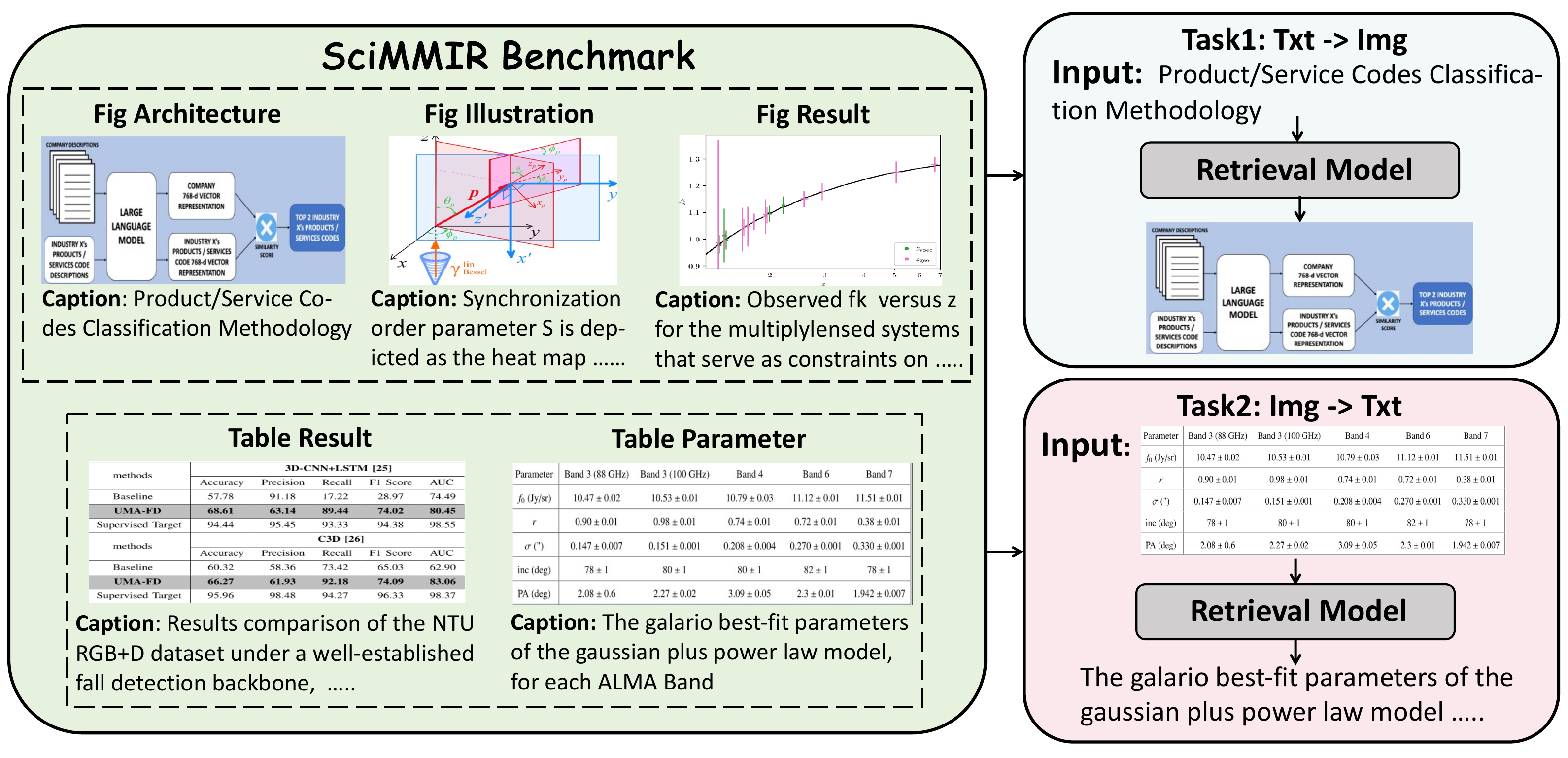}
    \caption{An illustration of the SciMMIR framework.}
    \label{fig:framework illustration}
\end{figure*}

To address the aforementioned research gap, we introduce \textbf{SciMMIR}, a \textbf{S}cientific \textbf{M}ulti-\textbf{M}odal \textbf{I}nformation \textbf{R}etrieval benchmark. SciMMIR (outlined in \autoref{fig:framework illustration}) is the first benchmark to comprehensively evaluate a model's MMIR ability in the scientific domain.
To build our data collection, we retrieve images of figures and tables, and their associated captions, from scholarly documents available on arXiv, an open-access archival corpus, to construct image-text pairs.
In order to comprehensively evaluate the cross-modality aligned representations learned by models, our SciMMIR benchmark defines the retrieval task as \textit{bi-directional}, involving searching for the correct textual caption in a candidate pool from a given image (\baktask) and finding the corresponding figure or table image from a textual caption (\fwdtask). 

Given the disparity among various data types, we contend that achieving uniform model performance across diverse data formats is challenging.
For example, a model may excel at retrieving data related to experimental results but demonstrate average performance regarding data related to model architectures. If an overall improvement is sought for the performance of VLMs, this improvement may not be observed in specific sub-domains of information. Consequently, such improvements do not necessarily translate into observable boosts to a VLM's performance for a specific use case.
As a result, we annotate and categorise the image-text pairs into three figure-caption and two table-caption subcategories based on the type of content they describe (such as experimental results, model architectures, and scientific principles). We then conduct \textbf{\textit{fine-grained evaluation on each subset}}. By analysing performance across subcategories, we are better able to carry out targeted improvements to a model for a specific subcategory of interest.

To explore the MMIR capabilities of existing image captioning models and VLMs in scientific domains, as well as different subcategories, we conduct extensive experiments in both zero-shot and fine-tuned settings across various subcategories. Furthermore, we extract OCR-text data from the images and investigate its influence on the performance of VLMs. We present our key insights as follows:
\begin{enumerate}
\itemsep -1mm
    \item We reveal that MMIR tasks in the scientific domain pose significant challenges for current VLMs, which usually do not demonstrate adequate performance in scientific domains.
    Furthermore, after fine-tuning VLMs with data specific to scientific domains, there is a marked performance improvement, underlining the effectiveness of domain-specific adaptation.
    \item 
    The results suggest a distinction between tasks involving the figure and table subsets, with performance on the figure subset being more easily improved through domain-specific adaptation.
    Furthermore, by leveraging text data extracted through OCR, we are able to substantially boost the performance of VLMs in MMIR tasks within scientific domains. This suggests that character recognition is a key weakness of standalone VLMs in the performance of SciMMIR.
    \item Regardless of parameter size, the BLIP-2 series of models generally perform better on SciMMIR than other pre-trained VLMs. This improved zero-shot capability may be the result of distinct pre-training tasks including image-text matching and image-text contrastive learning, rather than standard language modelling.
\end{enumerate}
These findings underscore the importance of tailored approaches for different data types within the scientific MMIR framework. A more in-depth exploration of these findings is given in \S\ref{sec: result}.

\section{Related Work}

\paragraph{General Information Retrieval.}

Information Retrieval is a fundamental task within NLP and has recently been facilitated by dense representation learning \cite{reimers2019sentence,karpukhin2020dense}. More recently, the desire for unified representations across tasks has become significant, with this line of research proposing to understand and evaluate task-agnostic representations in a single representation space \cite{muennighoff2023mteb,asai2022task,su2022one,wei2023uniir}. In another vein, domain generalisation has always been seen as a key weakness of IR models \cite{thakur2021beir}. 
Through the subpar performance of general image-text models on SciMMIR, we evidence that scientific IR, especially when multi-modal, remains an out-of-domain (OOD) task despite advancements in general information retrieval.

\paragraph{Multi-modal Information Retrieval.} 

In earlier multi-modal representation learning research, small-scale cross-modal retrieval datasets including MSCOCO \cite{lin2014mscoco} and Flickr30k \cite{plummer2015flickr30k} have facilitated the alignment between visual and linguistic representations. Efforts have since shifted towards large-scale vision-language pretraining \cite{radford2021learning,kim2021vilt,li2021align,jia2021scaling,yu2022coca}, with these small-scale retrieval datasets, in turn, becoming the standard evaluation approach for such systems. 
Advancements in multi-modal representation alignment have also facilitated multi-modal retrieval-augmented generation \cite{chen2022murag,yasunaga2022retrieval,hu2023reveal,lin2023fine}, and
more recently, evaluating the unified cross-modal representations across diverse tasks has emerged as a prevalent trend \cite{wei2023uniir}.

\paragraph{Scientific Document Retrieval.} 

Scientific information retrieval has received moderate attention in NLP, with SciFact \cite{wadden2020fact} and SCIDOCS \cite{cohan2020specter} commonly incorporated in popular zero-shot information retrieval benchmarks \cite{thakur2021beir}. More complex tasks have been proposed in this area, such as DORIS-MAE, a task to retrieve documents in response to complex, multifaceted scientific queries \cite{wang2023scientific}.
In the multi-modal area, VQA \cite{Antol_2015_ICCV} presents another major approach in evaluating vision-language systems, concerning in-depth visual grounding, rather than the use of distributional priors \cite{agrawal2018don}. It is in this area that work with a similar scope to ours in the scientific domain, such as PlotQA \cite{methani2020plotqa} and ChartQA \cite{masry2022chartqa}, is seen.
Our proposed SciMMIR benchmark distinguishes itself from these existing works by offering extensive coverage across annotations of figure and table subcategories, a larger dataset size, and the use of real-world data that is naturally paired and therefore not reliant on costly human annotation.

\begin{table}[tb]\centering
\resizebox{0.99\columnwidth}{!}{
\begin{tabular}{llrrrc}\toprule
\multicolumn{1}{c}{\multirow{2}{*}{\textbf{Subset}}} & \multicolumn{1}{c}{\multirow{2}{*}{\textbf{Subcategory}}} &\multicolumn{3}{c}{\textbf{Number}}  &\multicolumn{1}{c}{\textbf{Len (words)}} \\\cmidrule{3-6}
& &\textbf{Train} &\textbf{Valid} &\textbf{Test} &\multicolumn{1}{c}{\textbf{Caption}} \\\midrule
\multirow{3}{*}{\textbf{Figure}} & Result &296,191 &9,676 &9,488 & 52.89 \\
&Illustration &46,098 &1,504 &1,536 & 38.44 \\
&Architecture &13,135 &447 &467 & 27.27 \\ \midrule
\multirow{2}{*}{\textbf{Table}} & Result &126,999 &4,254 &4,229 & 27.23\\
& Parameter &15,856 &552 &543 & 17.10 \\ \midrule
\multicolumn{2}{c}{Total} &498,279 &16,433 & 16,263 & 43.19 \\ %
\bottomrule
\end{tabular}
}
\caption{Statistics of the SciMMIR dataset. 
}
\label{tab: Statistics of the SciMMIR}
\end{table}
\section{Dataset Construction}

\begin{figure}[tp]
\centering
\includegraphics[width=0.75\columnwidth]{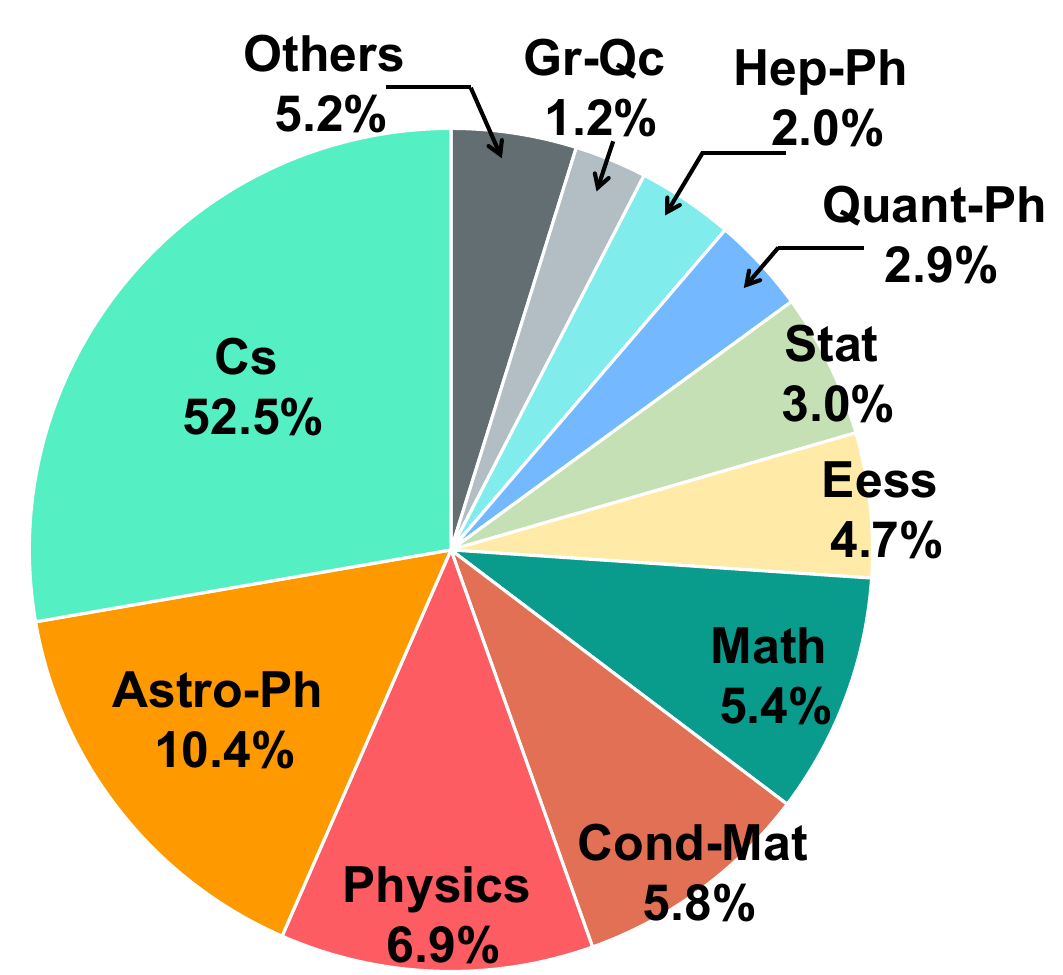}
\caption{The ratio of different subject image-caption data in SciMMIR. 'Hep-Ph' denotes High Energy Physics - Theory, 'Quant-Ph' denotes Quantum Physics, and 'Astro-Ph' denotes Astrophysics. 'Gr-Qc' denotes General Relativity and Quantum Cosmology, 'Eess' denotes Electrical Engineering and Systems Science, 'Cond-Mat' denotes Condensed Matter, and 'Cs' denotes Computer Science.}
\label{figure: Arxiv Statics}
\end{figure}

\paragraph{Data Collection.}
We collect PDF files from a 6-month period (i.e. papers submitted between May and October 2023) from arXiv using the official API\footnote{\url{https://info.arxiv.org/help/api}}.
We use an open-source tool~\cite{pdffigures2} to locate non-textual elements (i.e., figures and tables) in the papers and extract their corresponding caption text.
All tables and figures are stored in the form of images, and we remove the figure/table entries that have empty captions.
The aforementioned collection process results in the SciMMIR dataset that comprises 530K image-caption samples, with an average caption length of $43.19$ words as shown in \autoref{tab: Statistics of the SciMMIR}. 
The dataset is split into training, validation, and testing sets with $498,279$, $16,433$, and $16,263$ samples, respectively.
As shown in \autoref{figure: Arxiv Statics}, the SciMMIR benchmark covers a wide range of scientific disciplines, including those that require complex reasoning (such as Mathematics, Physics, and Computer Science), which attests to the presence of comprehensive and intricate scientific knowledge within the dataset.

\paragraph{Subset and Subcategory Structure.} 
To better understand the performance of VLMs across various data types within the scientific domain, we define a hierarchical architecture with \textit{two subsets} and \textit{five subcategories} for the SciMMIR benchmark.
We divide the data into two subsets: one for tables and one for figures, as they possess distinct data distributions. Tables contain ample textual information, whereas figures predominantly utilise geometric shapes to elucidate scientific principles or reveal patterns within data.
For tabular data, we further categorise them into two subcategories, \textit{Table-Parameter} and \textit{Table-Result}.  Table-Result data primarily presents experimental results, whereas Table-Parameter data provides explanations of parameters or specific numerical values (i.e., learning rates and physical coefficients), and consequently both have different data type distributions.
As for Figures, we consider those depicting experimental results, explaining model architectures (e.g. a figure describing each module in a deep learning model), and illustrating scientific theories (e.g. a figure illustrating an event related to double-slit interference, aiming to elucidate the underlying scientific principle), as they encompass different types of scientific knowledge. Therefore, the performance of models on these distinct data types may vary, leading us to categorise them into three separate subcategories. Specifically, our fine-grained categorisation is derived based on the statistics in  \autoref{tab:classification_fig_table}. 

\begin{table}[tb]\centering
\resizebox{0.99\columnwidth}{!}
{
\begin{tabular}{ll|p{0.6\columnwidth}}
\toprule
\centering \textbf{Subset} & \multicolumn{1}{l}{\textbf{Subcategory}} & \multicolumn{1}{c}{\textbf{Description}} \\\midrule
\multirow{6.5}{*}{\textbf{Figure}} 
& \multirow{2}{*}{Architecture} & Depicts scientific study frameworks and conceptual designs. \\\cmidrule{2-3}
& \multirow{2}{*}{Illustration} & Illustrates complex scientific concepts or data relationships. \\\cmidrule{2-3}
& \multirow{2}{*}{Result} & Visually presents scientific research outcomes. \\

\midrule

\multirow{4.5}{*}{\textbf{Table}}
& \multirow{2}{*}{Parameter} & Details of key parameters and variables in studies. \\\cmidrule{2-3}
& \multirow{2}{*}{Result} & Summarises and displays experiment/study results. \\

\bottomrule
\end{tabular}
}
\caption{The hierarchical structure of SciMMIR.}
\label{tab:classification_fig_table}
\end{table}

\paragraph{Data Annotation.} 
For data annotation, we use manually constructed key phrases to classify image-text sample pairs. Firstly, we acquire keywords based on the unique words that emerge in captions under different subcategories and conduct an initial categorisation of the data based on this keyword set. Subsequently, to ensure the quality of our statistical analysis, we randomly select 2000 images from the test set and hire three graduate students experienced in natural language processing to \textit{manually review} the results of the keyword-based classification on the criteria of whether the image within the image-caption pairs conform to the expected characteristic of the corresponding subcategory. We then construct new keywords and remove low-quality ones by analysing which words in the caption result in misclassified examples.
Finally, we refine the keyword list iteratively, enhancing the quality until the manual evaluation's accuracy on the 2,000 extracted samples reached 80\%.
The subset and subcategory classification results are shown in ~\autoref{tab: Statistics of the SciMMIR}, providing a structured and standardised basis for subsequent experiments.

\begin{table*}[!hbtp]%
\centering
\scriptsize
\resizebox{2.0\columnwidth}{!}{
\begin{tabular}{l|p{0.5\columnwidth}|r|p{0.5\columnwidth}|r|r|r}\toprule
\multirow{2}{*}{\textbf{Model}} &\multicolumn{2}{c}{\textbf{Pre-training Data}} & \multicolumn{1}{|c|}{\multirow{2}{*}{\textbf{Pre-training Task}}} & \multicolumn{3}{c}{\textbf{Trainable \& *{Frozen} Parameters}} \\
 & \multicolumn{1}{l}{\textbf{Domain}} &\multicolumn{1}{c|}{\textbf{Number}} & &\textbf{Visual} &\textbf{Textual} & \textbf{Align} \\\midrule\midrule
CLIP-base & \underline{Internet Crawled} &400M &Contrastive &62M &63M &/ \\\midrule
\multirow{2}{*}{BLIP-base} & COCO, VG, CC3M, CC12M, SBU, LAION-400M &\multirow{2}{*}{129M}& Image-Text Contrastive, Image-Text Matching, Language Modeling & \multirow{2}{*}{25.5M} & \multirow{2}{*}{108M} & \multirow{2}{*}{/} \\\midrule
BLIP2-OPT-2.7B & \multicolumn{1}{l|}{\multirow{6}{0.3\columnwidth}{COCO, VG, CC3M, CC12M, SBU, LAION-400M}} &\multirow{6}{*}{129M} &\multirow{6}{0.3\columnwidth}{Image-Text Contrastive, Image-Text Matching, Image-grounded Text Generation} &\multirow{6}{*}{*{1.3B}} &*{2.7B} &*{2.7B} \\ \cmidrule{1-1}\cmidrule{6-7}
BLIP2-OPT-6.7B & & & & &*{6.7B} &*{6.7B} \\\cmidrule{1-1}\cmidrule{6-7}
BLIP2-FLAN-T5-XL & & & & &*{2.85B} &*{2.85B} \\\cmidrule{1-1}\cmidrule{6-7}
BLIP2-FLAN-T5-XXL & & & & &*{11.3B} &*{11.3B} \\ \midrule
\multirow{2}{*}{LLaMA-Adapter2-7B} & {LAION-400M, COYO, MMC4, SBU, CC3M, COCO} &\multirow{2}{*}{56.7M} & \multirow{2}{*}{Fine-Tuning only} & \multirow{2}{*}{*{62M}} &\multirow{2}{*}{*{7B}} & \multirow{2}{*}{14M} \\ \midrule
Kosmos-2 & \underline{GRIT} &90M &Language Modeling &0.3B &1.3B &19M \\ \midrule
\multirow{2}{*}{mPLUGw-OWL2} & {COCO, CC3M, CC12M, LAION-5B, COYO, DataComp} & \multirow{2}{*}{400M} & \multirow{2}{*}{Language Modeling} & \multirow{2}{*}{0.3B} & \multirow{2}{*}{7B} & \multirow{2}{*}{0.9B} \\ \midrule
LLaVA-V1.5-7B & LAION, CC, SBU, ShareGPT  & 392M & Language Modelling & 0.3B& 6.9B & 0.02B \\
\bottomrule
\end{tabular}
}
\caption{The pre-training information of the baselines. "\_" refers to non-public or not fully public data.}\label{tab: model_info}
\end{table*}

\section{Experiment}

\subsection{Retrieval Baseline} 

To investigate the capabilities of current VLMs on the SciMMIR task and to assess whether data from different categories influences their performance, we evaluate a wide range of baseline models. Furthermore, we collect information regarding the pre-training strategy for each baseline model in ~\autoref{tab: model_info} and present additional details in Appendix~\ref{sec:pt_dataset}, in order to explore the potential factors that cause performance differences between VLMs.

\paragraph{Image Captioning Models.} 
As our baselines, we present image-captioning models, including \textbf{CLIP-base}~\cite{radford2021learning} and \textbf{BLIP-base}~\cite{li2022blip}, that have learned the pairing relationship between images and the corresponding text via a strong supervision signal. 
We evaluate these image captioning models trained on general domain datasets (such as images related to scenery and daily life events) in both zero-shot and fine-tuned settings to investigate the need for scientific domain adaptation. 
We also introduce \textbf{BERT}~\cite{devlin2018bert} as an alternative text encoder for captioning (denoted "+BERT" in the tables), where such ensemble baselines may reveal the influence of the text encoders.

\paragraph{Visual Language Models.} 
Additionally, we select large visual language models (VLMs) trained for multi-modal tasks such as Visual Question Answering (VQA) to examine their zero-shot and fine-tuned MMIR performance in the scientific domain. Additional details of the benchmarked VLMs are given in Appendix~\ref{sec:VLMs}.

\paragraph{OCR Based Method.} We perform OCR on the images in our SciMMIR benchmark to extract textual content. To improve the performance of VLMs on the SciMMIR task, we combine the OCR text embeddings generated by the text encoder of the VLMs with the image embeddings produced by the VLMs' visual encoder.

\subsection{Evaluation Protocol}

\paragraph{Task Definition.} 
The SciMMIR benchmark presents a bi-directional MMIR task:
\begin{itemize}
\itemsep -1mm
    \item \textbf{\fwdtask}: The forward direction retrieval task, where for a given text, the model retrieves the correct corresponding image from a candidate set.
    \item  \textbf{\baktask}: The inverse direction retrieval task, where given an image, the model retrieves the correct corresponding text from a candidate set.
\end{itemize}

Given an image $img_i$ and a text $text_j$, the relevance score $R$ in the retrieval ranking is defined as the dot product between the visual and textual representations of $img_i$ and $text_j$ (i.e.  
$R=E_{img_i} \cdot E_{text_j}).$
In addition to assessing the models' performance on the overall test set (denoted ``ALL''), we evaluate the models' retrieval capability on different subsets and subcategories to scrutinise their abilities. Specifically, we assess the models' performance on five fine-grained subcategories (shown in Table \ref{tab:classification_fig_table}) of the test set,
as well as the performance on the Figure and Table subsets overall.

\paragraph{Metrics.}

In this paper, we use the Mean Reciprocal Rank (MRR) and Hits@K metrics to assess the IR models' performance on the SciMMIR benchmark. These metrics are calculated based on the ranking of the golden answer within the entire set of candidates provided by the IR models. The details of these metrics are described in Appendix~\ref{metric}.

\paragraph{Zero-shot} 
We provide a zero-shot (ZS) setting in the evaluation for all baselines.
For \textit{image-captioning} models, the features extracted by the visual encoder and textual encoder are directly used, since they have been aligned to the same representation space. 
For the \textit{visual language} models, the visual representation remains unchanged, but the representations from the textual module are used depending on their architectures.
Specifically, for the encoder-decoder textual models such as BLIP2-FLAN-T5s, we use the output features from the textual encoder as the text features, 
whereas for decoder-only textual models like BLIP2-OPTs, we perform mean pooling on the outputs from the last decoder layer.

\paragraph{Fine-tuning.} 
We also provide an evaluation of fine-tuned (FT) versions of the relatively small models (CLIP-base and BLIP-base) and a large VLM (BLIP2-FLAN-T5-XL) that were trained with our data. 
During fine-tuning, we employ standard contrastive learning~\cite{chen2020constrastive} to maximize the relevance score between positive text-image pairs and minimise the relevance score between negative text-image pairs within a batch of samples. 
In addition to training the models on the entire training set, we also train them on different subsets (e.g., Figure-Result and Table-Parameter) of the training data to investigate the modelling abilities in a fine-grained manner.

\begin{table*}[hbt!]%
\begin{center} 
\footnotesize
\resizebox{2.0\columnwidth}{!}{
    \begin{tabular}{l|l|cccc|rrrr|rrrr}
\toprule
\multicolumn{1}{l|}{}                                        &                                        & \multicolumn{4}{c|}{\textbf{ALL}}                                                                              & \multicolumn{4}{c|}{\textbf{Figure*}}                                                                           & \multicolumn{4}{c}{\textbf{Table*}}                                                                            \\
\multicolumn{1}{l|}{}                                        &                                        & \multicolumn{2}{c}{\fwdtask} & \multicolumn{2}{c|}{\baktask} & \multicolumn{2}{c}{\fwdtask} & \multicolumn{2}{c|}{\baktask} & \multicolumn{2}{c}{\fwdtask} & \multicolumn{2}{c}{}{\baktask} \\
\multicolumn{1}{l|}{\multirow{-3}{*}{}}                      & \multirow{-3}{*}{\textbf{Model}}       & MRR                        & Hits@10                   & MRR                       & Hits@10                     & \multicolumn{1}{c}{MRR}  & \multicolumn{1}{c}{Hits@10} & \multicolumn{1}{c}{MRR}  & \multicolumn{1}{c|}{Hits@10} & \multicolumn{1}{c}{MRR}  & \multicolumn{1}{c}{Hits@10} & \multicolumn{1}{c}{MRR}  & \multicolumn{1}{c}{Hits@10} \\ 
\midrule\midrule
     & CLIP-base &\ \ 8.13	&13.48	& \ \ 7.94	&13.34	&\ \ 9.29	&15.41	&\ \ 8.99	&15.29	&\ \ 5.29	&\ \ 8.82	&\ \ 5.41	&\ \ 8.65                    \\
     & CLIP-base+BERT           & \ \ 2.47	& \ \ 5.01	& \ \ 3.11	& \ \ 5.85	&\ \ 2.99	&\ \ 6.09	&\ \ 3.80	&\ \ 7.10	&\ \ 1.19	&\ \ 2.42	&\ \ 1.44	&\ \ 2.85                     \\
     & BLIP-base                   & \ \ 6.14	& 11.30	& \ \ 6.18	&11.71	& \ \ 6.80	&12.59	&\ \ 6.89	&13.21 &\ \ 4.59	&\ \ 8.22	&\ \ 4.47	&\ \ 8.15 \\
     & BLIP-base+BERT                   & 11.51	&20.09	&12.69	&21.77 &{13.01}	&22.67 &14.12	&24.18	&\ \ 7.93	&13.98	&\ \ 9.31	&16.08 \\
     
     & BLIP2-FLAN-T5-XL &\ \ 4.44 	&\ \ 7.74 	&\ \ 2.27 &\ \ 4.48  &\ \ 4.93 	&\ \ 8.66 	&\ \ 2.57 	&\ \ 5.02 	&\ \ 3.23 	&\ \ 5.48 	&\ \ 1.51 	&\ \ 3.13 \\

\multirow{-5}{*}{\textbf{FT}}                       & CLIP-base+OCR                  & \textbf{20.23}	& \textbf{29.60}	& \textbf{20.70	}& \textbf{30.19}	& \textbf{20.38} 	& \textbf{29.71} 	& \textbf{20.87 }	& \textbf{30.49} 	& \textbf{20.00} 	& \textbf{29.49} 	& \textbf{20.41} 	& \textbf{29.60} 
          \\ 
\midrule
    & CLIP-base    &0.419	&0.719	&0.364	&0.670	&0.458	&0.767	&0.421	&0.787	&0.310	&0.586	&0.219	&0.375\\
   & BLIP-base   &0.004	&0.006	&0.003	&0.006	&0.006	&0.009	&0.002	&0.000	&0.001	&0.000	&0.007	&0.021\\
   & BLIP2-FLAN-T5-XL &0.025	&0.031	&0.012	&0.025	&0.028	&0.035	&0.016	&0.035	&0.020	&0.021	&0.003	&0.000\\
   & BLIP2-FLAN-T5-XXL & 0.053	&0.105	&0.004	&0.000	&0.059	&0.104	&0.004	&0.000	&0.040	&0.105	&0.003	&0.000 \\
    & BLIP2-OPT-2.7B  & 0.052	&0.111	&0.015	&0.031	&0.035	&0.060	&0.013	&0.027	&0.093	&0.230	&0.020	&0.042\\
    & BLIP2-OPT-6.7B  & 0.002	&0.006	&0.002	&0.000	&0.003	&0.008	&0.002	&0.000	&0.002	&0.000	&0.002	&0.000 \\
 & LLaVA-V1.5-7B &0.006 	&0.012 	&0.002 	&0.000 &0.008 	&0.018 	&0.002 	&0.000 	&0.002 	&0.000 	&0.002 	&0.000  \\
  & mPLUG-Owl2-LLaMA2-7B &0.002	&0.000	&0.002	&0.000	&0.003	&0.000	&0.002	&0.000	&0.001	&0.000	&0.001	&0.000\\
  & Kosmos-2 & 0.008	&0.018	&0.002&	0.000	&0.011	&0.025	&0.002	&0.000	&0.000	&0.000	&0.001	&0.000  \\
    \multirow{-10}{*}{\textbf{ZS}} & LLaMA-Adapter2-7B &0.040	&0.061	&0.002	&0.000	&0.056	&0.085	&0.002	&0.000	&0.001	&0.000	&0.004	&0.000 \\
\bottomrule
\end{tabular}
}
\end{center}
\caption{
The main results of SciMMIR benchmark.
* refers to average results in the Figure and Table subsets.
}
\label{tab:Main Result}
\end{table*}
\section{Result Analysis}\label{sec: result}

\subsection{Overall Evaluation}
Following the evaluation protocol shown in \autoref{tab:Main Result},
we report the baseline performances on the universal set (ALL), Figure set, and Table set.

For both the forward ({\fwdtask}) and inverse ({\baktask}) tasks, we find that small models (e.g. CLIP and BLIP) fine-tuned with our in-domain scientific image-text data generally demonstrate superior performance in all settings of the SciMMIR benchmark. Specifically, the MRR of fine-tuned CLIP and BLIP models are over 6\% in all settings.
This underscores the necessity of domain adaption for improvement in the SciMMIR task. Our designed tasks remain challenging for most of the models. 
For tasks across both directions, the zero-shot capabilities of most large VLMs demonstrate relatively poor performance, with both the MRR and Hits@10 metrics falling below 0.23\% in the ALL setting.
It is worth mentioning that the CLIP-base model's zero-shot performance is the best across all VLMs with its MMR being over 0.3\%, which suggests CLIP maybe encounter some image-caption pair related to the scientific domain during the pre-training.

The performance of the fine-tuned multi-modal models in information retrieval involving both figures and tables is promising overall. 
However, in the non-OCR setting, the performance of the models is significantly higher on the Figure subset than on the Table subset, suggesting that the table retrieval task is more challenging. Conversely, when fine-tuning with the OCR-text data, there is not an explicit gap between the models' performance on the Figure and Table subsets.
We hypothesise that the lower performance on the table subset without OCR-text data may be due to the scarcity of table-style images in the pre-training datasets and the lack of textual perception ability in the visual encoders. 

Experimental results based on our SciMMIR benchmark demonstrate the limitations of existing VLMs for MMIR in scientific domains. However, by employing the high-quality data of SciMMIR for fine-tuning, the performance of VLMs can be effectively improved. Additionally, our experiments show that retrieving visual tables is challenging and requires thoroughly mining the semantic relations between caption information and textual data within tables.

\subsection{Zero-Shot Analysis}

To provide a more thorough analysis, we present the zero-shot performance of the baselines across different subcategories in ~\autoref{tab:zero-shot results of figure} and ~\autoref{tab:zero-shot results of table} in Appendix \ref{Appendix Zero-shot Analysis}. 

\begin{table*}[htb!]
\begin{center} 
\footnotesize
\resizebox{2.0\columnwidth}{!}{
    \begin{tabular}{l|l|rrrr|rrrr|rrrrr}\toprule
    \multirow{3}{*}{\textbf{Model}} &\multirow{3}{*}{\textbf{Training Data}} &\multicolumn{4}{c|}{\textbf{Fig Architecture}} &\multicolumn{4}{c|}{\textbf{Fig Illustration}} &\multicolumn{4}{c}{\textbf{Fig Result}} \\
    & &\multicolumn{2}{c}{\fwdtask} &\multicolumn{2}{c|}{\baktask} &\multicolumn{2}{c}{\fwdtask} &\multicolumn{2}{c|}{\baktask} &\multicolumn{2}{c}{\fwdtask} &\multicolumn{2}{c}{\baktask} \\
    & &MRR &Hits@10 &MRR &Hits@10 &MRR &Hits@10 &MRR &Hits@10 &MRR &Hits@10 &MRR &Hits@10 \\\midrule\midrule
    \multirow{6}{*}{CLIP-base} &All & 9.77 	&16.92 	&9.84 	&15.42 	&10.01 	&15.30 	&9.35 	&14.97 	&9.16	&15.37	&8.90 	&15.34 
 \\
    &Fig-Architecture &5.60 	&8.35 	&6.11 	&8.14 	&2.61 	&4.95 	&2.95 	&5.01 	&2.50 	&4.02 	&2.35 	&4.18 
 
 \\
    &Fig-Illustration & 8.58 	&12.85 	&8.82 	&13.28 	&6.76 	&11.72 	&7.08 	&11.78 	&5.69 &	9.20 	&5.46 	&8.96 
\\
    &Fig-Result &9.24 	&15.42 	&9.76 	&14.99	&8.58 	&14.19	&8.86 	&14.26 	&8.79 	&14.10 	&9.05 	&14.79 
\\
    &Table-Parameter & 2.67	&4.50 	&3.04 	&3.85 	&1.78 	&3.19 	&2.42 	&4.49 	&1.82 	&2.99 	&1.55 	&2.74 
\\
    &Table-Result & 3.12 	&5.78 &	3.31 	&5.35 &	1.91 	&3.91 	&2.33 	&4.49 &	2.58 &	4.26 &	1.48 &	2.80 
\\ \midrule
    CLIP-base+BERT &All & 2.30 	&4.93 	&2.76 &6.42 	&3.12 	&5.53 	&3.59 	&6.97 &	3.01 &	6.23 	&3.88 &	7.16 
\\ \midrule
    CLIP-base+OCR &All & 15.40 &	22.70 &	16.41 	& 25.48 &	15.89 	&23.24 	&16.94 	&24.61 &	21.29 &	31.03 	&21.68 	&31.63 

\\ \midrule
    \multirow{6}{*}{BLIP-base} &All & 5.11	&10.06 	&5.53 	&10.28 	&5.35 &	10.09 	&5.64 	&10.16 	&7.11 	&13.10 	&7.15 	&13.82
\\
    &Fig-Architecture &0.04	&0.00 	&0.06 	&0.21 	&0.02 &	0.00 &	0.03 	&0.07 	&0.03 &	0.06 &	0.02 &	0.01 
\\
    &Fig-Illustration & 0.04 &	0.00 	&0.09 &	0.00& 	0.26 	&0.52 	&0.45 	&0.91 	&0.08 &	0.16 &	0.09 &	0.14  
\\
    &Fig Result &2.55 	&6.21 	&3.20 	&6.00 	&2.91 	&6.25 &	3.380 	&6.84 	&4.66 &	9.13 &	4.80 &	9.18  
\\
    &Table-Parameter & 0.00 &	0.00 &	0.00& 	0.00 	&0.00 	&0.00 	&0.00 &	0.00 &	0.00 &	0.00& 	0.01 &	0.00   
\\
    &Table-Result &0.12 	&0.21 	&0.01 &	0.00 &	0.01 	&0.00 	&0.03 	&0.07 	&0.05 	&0.07& 	0.06 &	0.09 
\\ \midrule
    BLIP-base+BERT &All &9.95 &	18.42 & 	12.09 &	18.63 	&11.17 	&19.27 &	11.63 	&20.25 	&13.44 	&23.39 	&14.60 	&25.04 &
\\ \midrule
    BLIP2-FLAN-T5-XL &All &6.75 	&11.34 	&4.06 	&8.56 	&5.99 	&10.41 	&3.16 &	6.44 	&4.69 &8.27 	&2.41 	&4.64 
 \\
    \bottomrule
    \end{tabular}
}
\end{center}
\caption{The results of fine-tuning models on Figure subsets of our SciMMIR benchmark.
}
\label{tab:Fine-tuning models figure results}
\end{table*}
\begin{table*}[htb!]
\begin{center} 
\footnotesize
\resizebox{1.5\columnwidth}{!}{
    \begin{tabular}{l|l|rrrr|rrrrr}\toprule
    \multirow{3}{*}{\textbf{Model}} &\multirow{3}{*}{\textbf{Training Data}} &\multicolumn{4}{c|}{\textbf{Table Result}} &\multicolumn{4}{c}{\textbf{Table Parameter}} \\
    & &\multicolumn{2}{c}{\fwdtask} &\multicolumn{2}{c|}{\baktask} &\multicolumn{2}{c}{\fwdtask} &\multicolumn{2}{c}{\baktask} \\
    & &MRR &Hits@10 &MRR &Hits@10 &MRR &Hits@10 &MRR &Hits@10 \\\midrule\midrule
    \multirow{6}{*}{CLIP-base} &All & 5.40 	&9.01 	&5.52 	&8.82 	&4.45 	&7.37 	&4.55 	&7.37 

 \\
    &Fig-Architecture & 1.22 	&2.06 	&1.34 	&2.34 	&1.35 	&2.58 	&1.47 	&2.95  \\
    &Fig-Illustration & 1.42	&2.70 	&1.79 	&3.14 	&1.93 	&2.95 	&2.60 	&4.42  \\
    &Fig-Result &2.71 	&4.49 	&2.53 	&4.52 	&2.19 	&4.05 	&2.30 	&4.79
 \\
    &Table-Parameter & 1.46 	&2.70 	&1.56 	&2.62 	&1.52 	&3.31 	&1.82 	&3.68 
\\
    &Table-Result &4.28 	&7.26 	&1.28 	&2.29 	&3.77 	&6.63 	&0.87 	&1.29 
\\ \midrule
    CLIP-base+BERT &All &1.18 	&2.41 	&1.46 	&2.93 	&1.31 	&2.58 	&1.33 	&2.21 
\\ \midrule
    CLIP-base+OCR &All &20.36 	& 29.87 	& 20.68 	& 29.96 	& 17.15 	& 26.52 	& 18.22 	& 26.70 

\\ \midrule
    \multirow{6}{*}{BLIP-base} &All & 4.77 	&8.42 	&4.54 	&8.23 	&3.16 	&6.63 	&3.99 	&7.55 
\\
    &Fig-Architecture &0.01 	&0.00 	&0.03 	&0.02 	&0.01 	&0.00 	&0.02 	&0.00 
\\
    &Fig-Illustration &0.00 	&0.00 	&0.01 	&0.00 	&0.01 	&0.00 	&0.02 	&0.00 
\\
    &Fig-Result &0.70 	&1.32 	&0.65 	&1.16 	&0.32 	&1.29 	&0.56 	&0.74 
\\
    &Table-Parameter &0.01	&0.02 	&0.01 	&0.00 	&0.02	&0.00 	&0.06 	&0.00 
 \\
    &Table-Result &0.92	&1.80 	&0.92 	&1.82 	&0.83 	&0.74 	&0.52 	&1.10 
\\ \midrule
    BLIP-base+BERT &All &8.17 	&14.35 	&9.70 	&16.48 	&6.01 	&11.05	&6.19	&12.89
\\ \midrule
    BLIP2-FLAN-T5-XL &All &3.11 	&5.29 	&1.33 	&2.90 	&4.22 	&6.99 &3.00 	&4.97 
 \\
    \bottomrule
    \end{tabular}
}
\end{center}
\caption{The results of fine-tuning models on Table subsets of our SciMMIR benchmark.
}
\label{tab:Fine-tuning models table results}
\end{table*}

As for the \fwdtask~direction, the selected large pre-trained VLMs (e.g. BLIP2-OPT-6.7B and LLaMA-Adapter2-7B) demonstrate poor performance across various subcategories in both the Figure and Table subsets.
For the subcategories within the Table subset, all models, except CLIP-base, are ineffective.
In the Figure subset, the BLIP2-FLAN-T5 series models show slightly better performance across all subcategories. This may be attributed to the fact that the textual encoder part of encoder-decoder architecture could better capture textual features.
Conversely, as for the \baktask~ direction, on the Figure subset, the performance of all VLMs in the reverse direction is marginally lower than in the forward direction.
This proves that the image feature captured by visual encoder of current VLMs is unable to model effective relation with the relevant text.

\subsection{Analysis in Fine-tuning Setting}

\paragraph{Overall Analysis.}

As shown in ~\autoref{tab: fine-tuning model on different training data} in Appendix \ref{Appendix Fine-tuning Analysis}, we fine-tune the models using data of different categories.
The results indicate that training the model only with data from a specific subcategory leads to a significant performance gap compared to the models fine-tuned on all the data.
There are two main factors contributing to this. Firstly, the dataset size of a specific subcategory is relatively small. Secondly, there are significant differences in data distributions among different subcategories. 

Among all the models, BLIP-base+BERT performs the best across all fine-tuned settings, while the performance of the CLIP model drops when its text encoder is replaced with BERT.
Notably, merely fine-tuning the Q-Former parameters of BLIP2-FLAN-T5-XL to adapt the large VLM to the scientific domain did not yield as effective results as the smaller models.
Consequently, there remains a need for efficiently fine-tuning small models to construct robust connections between the representations of the visual and textual modalities.

\begin{table*}[!htp]
\centering
\label{tab: results of fine-tuning models of all}
\small
\resizebox{2.0\columnwidth}{!}{
    \begin{tabular}{l|l|rr|rr|rr|rr|rrr}\toprule
    \multirow{2}{*}{\textbf{Model}} &\multirow{2}{*}{\textbf{Testing Data}} &\multicolumn{2}{c|}{\textbf{Fig-Architecture}} &\multicolumn{2}{c|}{\textbf{Fig-Illustration}} &\multicolumn{2}{c|}{\textbf{Fig-Result}} &\multicolumn{2}{c|}{\textbf{Table-Result}} &\multicolumn{2}{c}{\textbf{Table-Parameters}} \\
    & &\fwdtask &\baktask &\fwdtask &\baktask &\fwdtask &\baktask &\fwdtask &\baktask &\fwdtask &\baktask \\\midrule\midrule
    \multirow{6}{*}{\textbf{FT-CLIP-base}} &Fig Architecture &12.85	&12.72	&16.62	&18.22	&69.57	&67.22	&0.84	&1.65	&0.13	&0.19 \\
    &Fig Illustration &5.16	&4.66	&20.59	&22.66	&73.30	&71.47	&0.83	&0.98	&0.13	&0.23 \\
    &Fig Results &3.80	&3.62	&13.01	&14.25	&81.48	&80.15	&1.48	&1.64	&0.22	&0.34 \\
    &Table Results &0.12	&0.15	&0.24	&0.70	&4.16	&4.97	&85.68	&84.29	&9.81	&9.89\\
    &Table Parameters &0.29	&0.35	&0.53	&1.34	&5.08	&9.61	&73.44	&72.19	&20.64	&16.50\\
    \midrule
    \multirow{6}{*}{\textbf{ZS-CLIP-base}} &Fig Architecture &7.34	&6.72	&28.54	&23.06	&59.42	&66.62	&4.20	&2.70	&0.49	&0.90 \\
    &Fig Illustration &3.99	&3.68	&30.56	&23.44	&61.74	&71.04	&3.40	&1.47	&0.31	&0.36\\
    &Fig Results &4.12	&4.17	&24.31	&19.59	&63.04	&73.52	&7.74	&2.29	&0.79	&0.44 \\
    &Table Results &0.36	&2.55	&1.48	&4.91	&9.28	&38.69	&75.89	&41.92	&12.99	&11.92 \\
    &Table Parameters &0.26	&3.00	&2.38	&7.38	&9.52	&42.43	&74.40	&34.68	&13.44	&12.50 \\
    \bottomrule
    \end{tabular}
}
\caption{The accuracy and error analysis of CLIP models on our SciMMIR benchmark.}
\label{error analysis}
\end{table*}
\paragraph{The Impact of Subcategory Training Data.}

As shown in ~\autoref{tab:Fine-tuning models figure results} and \autoref{tab:Fine-tuning models table results}, we report the result on testing samples of specific subcategories, for the sake of comprehensively investigating the impact of different subcategory training data.

For BLIP, training on a certain subcategory results in performance improvements on the corresponding part of the test set, whilst its performance on other subcategories remains relatively poor. 
This demonstrates the distribution gap among our labelled subcategories and proves the rationality of our subset-subcategory hierarchy.
As for CLIP, the models trained on different subcategories consistently perform best in the Fig-Architecture subcategory. This may be because CLIP has been trained on data with a more similar distribution to our data.

The model trained on Figure-Results data demonstrates the best performance across the entire Figure subset. One reason could be that the Figure-Result subset has the largest training proportion (54.02\%) and contains text documents with a relatively longer average length (52.93 words compared to the dataset's overall average length of 43.23 words) in the training dataset.
This highlights the impact of training dataset size and its length coverage of text \cite{xiao2023length} on the performance and generalisability of retrieval models.

\paragraph{The Impact of OCR.}
After fine-tuning the CLIP model with OCR-extracted text data, we observe a notable improvement on the Table subset compared to the Figure subset. Furthermore, as for the subcategories related to results (i.e., the Table Result and Fig Result subcategories), the VLMs achieve their best performance, with MRR exceeding 20\%, compared to other models and subcategories.
These findings indicate that the OCR-extracted text data can provide textual information from the images, which may not be completely captured by the VLMs. This underscores the value of incorporating OCR data to enhance the text perception ability of VLMs, particularly in the domain of scientific multi-modal information retrieval.

\subsection{Text Encoder Generalisability}

To investigate the impact of text encoders on SciMMIR, we substitute the text encoders in both BLIP-base and CLIP-base models with BERT-base. 
As shown in ~\autoref{tab: fine-tuning model on different training data} in Appendix \ref{Appendix Fine-tuning Analysis}, replacing the text encoder of BLIP with BERT results in a significant improvement, while that of CLIP experiences a decline.
The reason for the performance changes after replacing the text encoder with BERT in both CLIP and BLIP may be as follows:

\paragraph{CLIP.}
With effective contrastive learning \cite{wang2020understanding}, the textual and visual embeddings are well-aligned in an isotropic space in the pre-training phase of CLIP, which is demonstrated by the zero-shot setting experiments.
However, replacing the text encoder with a highly anisotropic vanilla text encoder (e.g., BERT) hinders the stable alignment with the already learned vision encoder \cite{xiao2023isotropy}. We hypothesise that freezing the vision encoder in early fine-tuning may help guide the replaced language model.

\paragraph{BLIP.}
Unlike CLIP, BLIP incorporates BERT as its text encoder from the pre-training phase. This structural consistency significantly contributes to the model's enhanced adaptation to the scientific domain.
Besides, employing BERT as the text encoder may facilitate the learning of more effective text representations. This is particularly advantageous in establishing associations between images and text, especially since tables, a common element in scientific documents, are rich in textual information.

\subsection{Error Analysis}

For better analysis of the performance, we conduct experiments on test data spanning different subcategories and calculate the ratio of all subcategories within the top 10 answers predicted by the fine-tuned and vanilla CLIP models. Predictions matching the test subcategory were considered correct.

As shown in ~\autoref{error analysis}, due to the larger volume of samples in our dataset are labelled as Fig-Results and Table-Results (58.00\% and 26.16\%, calculated through Table \ref{tab: Statistics of the SciMMIR}, respectively), the models tend to predict samples from these categories as answers.
When comparing zero-shot and fine-tuned models, it can be observed that fine-tuning helps reduce
incorrect predictions across almost all categories. 

Compared with zero-shot results, the fine-tuned models show the largest improvement in prediction accuracy on the Figure-Architecture and Figure-Result test data. However, the increase in prediction accuracy on the Table subset after fine-tuning is not obvious, indicating that retrieving information from Tables still poses significant challenges.

\section{Conclusion}
In summary, we introduce a novel benchmark and a corresponding dataset designed to address the gap in evaluating multi-modal information retrieval (MMIR) models in the scientific domain. Additionally, we categorise the images into fine-grained subcategories based on the characteristics of the figures and tables to facilitate a more comprehensive evaluation and analysis. 
Our evaluation of zero-shot and fine-tuned approaches, which we conduct on extensive baselines within various subsets and subcategories, offers valuable insights for future research.

\section*{Limitations}
Due to computational resource constraints, we only fine-tune BLIP2-FLAN-T5-XL on our SciMMIR dataset and did not investigate the fine-tuning effects of other large VLMs on our benchmark.
In this work, we find that BLIP+BERT could improve the model's ability in our benchmark, specifically for the Table subset. However, we do not design experiments to explore which kind of models would be better suited to the replacement of the textual encoder with BERT or other language models.Despite our best efforts to ensure data quality, given the sheer volume of data, we cannot guarantee that all results and content within the scientific domain dataset are accurate. This inherent limitation could potentially lead to models generating misleading or deceptive outputs in future use, necessitating further filtering in future work.

\section*{Ethics Statement}
The dataset used in our research is constructed using publicly available data sources, ensuring that
there are no privacy concerns or violations. We do not collect any personally identifiable information, and all data used in our research is obtained
following legal and ethical standards.
In the stage of designing keywords and human evaluation classification of image-text pair, we employed three graduate students experienced in natural language processing for human evaluation. We paid the graduate students approximately \$13 per hour, well above the local average wage, and engaged in constructive discussions if they had concerns about the process.

\bibliography{custom}

\appendix

\section{The Baseline Pre-training Datasets}
\label{sec:pt_dataset}
Enhancing model performance through additional knowledge has garnered considerable attention \cite{cheng2023ml,cheng2023mrrl,cheng2023accelerating,wang2023cola} making it is essential to boost model capabilities in the scientific domain through multi-modal retrieval.
Besides, retrieving the related specific domain knowledge could significantly relieve the hallucination of LLM and VLMs \cite{choi2023kcts} and improve the interpretability of them \cite{wang2022subeventwriter,wang2024absinstruct}.
To this end, we have designed a series of baselines and provided a reference list for the pre-training image-text datasets mentioned in \autoref{tab: model_info}.
COCO~\cite{lin2014mscoco}, consists of over 200,000 images across various categories including people, animals, everyday objects, and indoor scenes. 
The VG dataset ~\cite{krishna2017vg} consists of over 100,000 images and covers a diverse range of visual concepts, including objects, scenes, relationships between objects, and other contextual information within images. 
CC3M~\cite{sharma2018cc3m} contains over 3.3  million of images paired with descriptive captions, covering a wide range of topics and scenes. 
CC12M~\cite{changpinyo2021cc12m} contains 12.4 million image-text pairs, which is 3 times larger in scale compared to CC3M with a higher diversity degree containing more instances of out-of-domain (OOD) visual concepts. 
SBU~\cite{ordonez2011sbu} contains over 1 million images with visually relevant captions. 
The dataset is designed to be large enough for reasonable image-based matches to a query and the captions are filtered to ensure they are visually descriptive and likely to refer to visual content. 
LAION-400M~\cite{schuhmann2021laion400m} is an open dataset that consists of 400 million image-text pairs, their CLIP embeddings, and KNN indices for efficient similarity search.  It includes image URLs, corresponding metadata, CLIP image embeddings, and various KNN indices for quick search. 
LAION-5B~\cite{schuhmann2022laion5b} is an open, large-scale dataset that consists of 5.85 billion image-text pairs, with 2.32 billion pairs in English. 
COYO~\cite{kakaobrain2022coyo-700m} is a large-scale dataset containing 747M image-text pairs as well as many other meta-attributes to increase the usability to train various models. 
MMC4~\cite{zhu2023mmc4} consists of 101.2 million documents with 571 million images interleaved with 43 billion English tokens. It covers a wide range of everyday topics such as cooking, travel, and technology. 
GRIT~\cite{peng2023kosmos} is a large-scale dataset of Grounded Image-Text pairs that consists of approximately 91 million images, 115 million text spans, and 137 million associated bounding boxes. 
DataCamp~\cite{gadre2023datacomp} is a participatory benchmark that focuses on dataset curation for large image-text datasets. It provides a new candidate pool of 12.8 billion image-text pairs. The dataset size in DataComp is a design choice and not predetermined. 

\section{Used Visual Language Models}
\label{sec:VLMs}
\begin{itemize}
\itemsep -1mm
    \item \textbf{BLIP-2}~\cite{li2023blip} series models use a querying transformer module to address the modality gap. We choose the models grounded in large language models (LLMs), BLIP2-OPT-2.7B, BLIP2-OPT-6.7B, BLIP2-FLAN-T5-XL and BLIP2-FLAN-T5-XXL, as our baselines.
    \item  \textbf{LLaVA-V1.5-7B}~\cite{liu2023improvedllava} use two simple methods, namely, an MLP cross-modal connector incorporating academic task related data such as VQA to improve the ability of the LLaVA.
    \item  \textbf{LLaMA-Adapter2-7B}~\cite{gao2023llamaadapterv2} efficiently fine-tunes additional parameters based on the LLaMA model~\cite{touvron2023llama}, where the extra expert models further boost its image understanding capability.
    \item \textbf{Kosmos-2}~\cite{peng2023kosmos} aligns perception with language and adds the ability to recognise and understand images based on its multi-turn dialogue and reasoning capabilities. Specifically, it achieves the capability of grounding images, allowing it to interact with inputs at the object level.
    \item \textbf{mPLUGw-OWL2}~\cite{DBLP:journals/corr/abs-2311-04257} introduces a Modality-Adaptive Module (MAM) into the large language model. By adding a small number of parameters during the attention process, it further learns a shared space for both vision and language representations.

\end{itemize}

\begin{table*}[t]
\begin{center} 
\footnotesize
\resizebox{2.0\columnwidth}{!}{
    \begin{tabular}{l|rrrr|rrrr|rrrr}
    \toprule
    \multirow{3}{*}{\textbf{Model}} & \multicolumn{4}{c|}{\textbf{Fig Architecture}}                     & \multicolumn{4}{c|}{\textbf{Fig Illustration}}                     & \multicolumn{4}{c}{\textbf{Fig Result}}                      \\
                           & \multicolumn{2}{c}{\fwdtask} & \multicolumn{2}{c|}{\baktask} & \multicolumn{2}{c}{\fwdtask} & \multicolumn{2}{c|}{\baktask} & \multicolumn{2}{c}{\fwdtask} & \multicolumn{2}{c}{\baktask} \\
                           & MRR         & Hits@10       & MRR          & Hits@10       & MRR          & Hits@10      & MRR          & Hits@10       & MRR         & Hits@10       & MRR         & Hits@10       \\ 
    \midrule\midrule
    CLIP-base             & 1.351 	&1.927 	&1.074 	&2.141 	&0.750 	&1.237 	&0.458 	&0.716 	&0.373 	&0.643 	&0.386 	&0.738 \\
    BLIP-base             & 0.003 	&0.000 	&0.001 	&0.000 	&0.003 	&0.000 	&0.002 	&0.000 	&0.006 	&0.011 	&0.002 	&0.000 \\
    BLIP2-FLAN-T5-XL       & 0.010 	&0.000 	&0.003 	&0.000 	&0.010 	&0.000 	&0.004 	&0.000 	&0.032 	&0.042 	&0.019 	&0.042 \\
    BLIP2-FLAN-T5-XLL      & 0.056 	&0.214 	&0.003 	&0.000 	&0.037 	&0.065 	&0.005 	&0.000 	&0.062 	&0.105 	&0.004 	&0.000 \\
    BLIP2-OPT-2.7B         & 0.130 	&0.214 	&0.005 	&0.000 	&0.033 	&0.130 	&0.006 	&0.000 	&0.031 	&0.042 	&0.014 	&0.032 \\
     BLIP2-OPT-6.7B         & 0.001 	&0.000 	&0.001 	&0.000 	&0.009 	&0.065 	&0.001 	&0.000 	&0.002 	&0.000 	&0.002 	&0.000 \\
    LLaVA-V1.5-7B  &0.003 	&0.000 	&0.004 	&0.000 	&0.003 	&0.000 	&0.004 	&0.000 &0.009 	&0.021 	&0.002 	&0.000 
 \\ 
    Kosmos-2 & 0.123 	&0.428 	&0.008 	&0.000 	&0.011 	&0.000 	&0.004 &0.000 	&0.006 	&0.011 	&0.002 	&0.000   \\
    mPLUG-Owl2-LLaMA2-7B &0.022 	&0.000 	&0.003 	&0.000 	&0.302 	&0.521 	&0.003 	&0.000 	&0.019 	&0.021 	&0.002 	&0.000  \\ 
    LLaMA-Adapter2-7B &0.001 	&0.000 	&0.001 	&0.000 	&0.008 	&0.000 	&0.002 	&0.000 	&0.002 	&0.000 	&0.002 	&0.000 \\
    \bottomrule
    \end{tabular}
}
\end{center}
\caption{The zero-shot results of multimodal models on Figure subsets of our SciMMIR benchmark.
}
\label{tab:zero-shot results of figure}
\end{table*}

\begin{table*}[t]
\begin{center} 
\footnotesize
\resizebox{1.6\columnwidth}{!}{
    \begin{tabular}{l|rrrr|rrrr}
    \toprule
    \multirow{3}{*}{\textbf{Model}} & \multicolumn{4}{c|}{\textbf{Table Result}}                                                                              & \multicolumn{4}{c}{\textbf{Table Parameter}}                                                                          \\
                           & \multicolumn{2}{c}{\fwdtask}                           & \multicolumn{2}{c|}{\baktask}                           & \multicolumn{2}{c}{\fwdtask}                           & \multicolumn{2}{c}{\baktask}                           \\
                           & \multicolumn{1}{c}{MRR} & \multicolumn{1}{c}{Hits@10} & \multicolumn{1}{c}{MRR} & \multicolumn{1}{c|}{Hits@10} & \multicolumn{1}{c}{MRR} & \multicolumn{1}{c}{Hits@10} & \multicolumn{1}{c}{MRR} & \multicolumn{1}{c}{Hits@10} \\ \midrule\midrule
    CLIP-base             &0.281 	&0.544 	&0.177 	&0.284 	&0.545 	&0.921 	&0.558 	&1.105 \\
    BLIP-base             & 0.001 	&0.000 	&0.007 	&0.024 	&0.000 	&0.000 	&0.003 	&0.000 \\
    BLIP2-FLAN-T5-XL       &0.021 	&0.024 	&0.003 	&0.000 	&0.010 	&0.000 	&0.005 	&0.000 \\
    BLIP2-FLAN-T5-XLL      & 0.041 	&0.095 	&0.003 	&0.000 	&0.030 	&0.184 	&0.003 	&0.000 \\
    BLIP2-OPT-2.7B         & 0.076 	&0.213 	&0.010 	&0.024 	&0.228 	&0.368 	&0.101 &0.184 \\
    BLIP2-OPT-6.7B         & 0.002 	&0.000 	&0.002 	&0.000 	&0.001 	&0.000 	&0.002 	&0.000 \\
    LLaVA-V1.5-7B                &0.002 	&0.000 	&0.002 	&0.000 &	0.003 	&0.000 &	0.004 	&0.000 
 \\ 
    Kosmos-2 & 0.000 	&0.000 	&0.001 	&0.000 &0.000 	&0.000 	&0.003 	&0.000  \\
    mPLUG-Owl2-LLaMA2-7B &0.001 	&0.000 	&0.004 	&0.000 	&0.002 	&0.000 	&0.005 	&0.000 \\
    LLaMA-Adapter2-7B &0.001 	&0.000 	&0.001 	&0.000 	&0.001 	&0.000 	&0.001 	&0.000 \\ \bottomrule
    \end{tabular}
}
\end{center}
\caption{The zero-shot results of multi-modal models on Table subsets of our SciMMIR benchmark datasets.
}
\label{tab:zero-shot results of table}
\end{table*}

\begin{table*}[!htb]
\centering
\scriptsize
\resizebox{1.4\columnwidth}{!}{
    \begin{tabular}{c|c|l|rr|rrr}\toprule
    \multirow{2}{*}{\textbf{Img Dim}} &\multirow{2}{*}{\textbf{Model}} &\multirow{2}{*}{\textbf{Training Dataset}} &\multicolumn{2}{c|}{\fwdtask} &\multicolumn{2}{c}{\baktask} \\
    & & &MRR &Hits@10 &MRR &Hits@10 \\\midrule\midrule
    \multirow{6}{*}{224} &\multirow{6}{*}{BLIP-base} &ALL & 0.958 	&2.034 	&1.138 	&2.294 	 
 
 \\
    & &Fig Architecture   &0.002 	&0.000 	&0.006 	&0.000 	

 \\
    & &Fig Illustration  &0.036 	&0.024 	&0.011 	&0.000

  \\
    & &Fig Result  & 0.167 &0.260 	&0.115 	&0.213 	

  \\
    & &Table Result  &0.408 	&0.757 	&0.368 	&0.686 

 \\
    & &Table Parameter  &0.011 	&0.024 	&0.009 	&0.000  

 \\ \midrule
    \multirow{1}{*}{224} &BLIP-base+BERT &ALL &1.614 	&3.334 	&2.102 	&4.375  
 \\ \midrule
    \multirow{6}{*}{384} &\multirow{6}{*}{BLIP-base} &ALL  &6.14	&11.3	&6.18	&11.71 \\
    & &Fig Architecture  &0.02	&0.04	&0.02	&0.02  \\
    & &Fig Illustration  &0.07	&0.14	&0.10	&0.17   \\
    & &Fig Result &3.26	&6.48	&3.40	&6.50  \\
    & &Table Result  &0.30	&0.54	&0.30	&0.57  \\
    & &Table Parameter  &0.01	&0.01	&0.01	&0.00  \\ \midrule
    \multirow{1}{*}{384}&BLIP-base+BERT &ALL  &11.51	&20.09	&12.69	&21.77 \\
    \bottomrule
    \end{tabular}
}
\caption{The averaged results of fine-tuning BLIP with different preprocessing image dimensions on \textit{ALL} testing candidates of our SciMMIR benchmark.}
\label{tab: fine-tuning BLIP with different preprocessing image dimensions}
\end{table*}
\section{Effects of Visual Encoder Resolution}
\label{sec:EVER}
In ~\autoref{tab:Main Result} for overall results, we compare the fine-tuned BLIP with the default image preprocessing dimensions of 384 and fine-tuned CLIP with the default image preprocessing dimensions of 224, where the results are relatively close. 
To make a fairer comparison, we decrease the image dimensions of BLIP-base model from 384 to 224 to be the same as CLIP-base to conduct SciMMIR evaluation, as described in \autoref{tab: fine-tuning BLIP with different preprocessing image dimensions}.

It can be seen that the granularity of image processing has a significant impact on model performance. When using a lower preprocessing dimension, the performance of BLIP is significantly decreased in both {\fwdtask} and {\baktask} tasks, using all training data settings.
The performance of the CLIP model, which uses the same image processing dimension, is almost double that of BLIP.

Furthermore, although replacing the text encoder of BLIP with BERT during training on lower-dimensional (224) image preprocessed data improved the performance of the model, there was still a significant gap compared to CLIP. However, when the text encoder of BLIP was replaced with BERT during training on higher-dimensional image preprocessed data, the performance of the model was far superior to both CLIP and CLIP+BERT.
This suggests that certain image-text shared interactive information is stored in the visual representations, and higher image quality can help the models better establish the connection between image and text representations.
\section{MRR and Hit@K}
\label{metric}
\begin{itemize}
\itemsep -1mm
    \item \textbf{MRR} stands for Mean Reciprocal Rank and is calculated as the reciprocal of the golden label's ranking in candidates. A higher MRR score indicates better performance.
    \item  \textbf{Hits@K} assesses the accuracy of the retrieval system by checking whether the golden label is present within the top-k ranked results. Hits@10 are used in our measurements.
\end{itemize}

\section{Fine-tuning Analysis}
\label{Appendix Fine-tuning Analysis}
\begin{table}[!htp]
\centering
\scriptsize
\resizebox{0.99\columnwidth}{!}{
    \begin{tabular}{l|l|rr|rr}\toprule
    \multirow{2}{*}{\textbf{Model }} &\multirow{2}{*}{\textbf{Training Dataset}} &\multicolumn{2}{c|}{\fwdtask} &\multicolumn{2}{c}{\baktask} \\
    & &MRR &Hits@10 &MRR &Hits@10 \\\midrule\midrule
    \multirow{6}{*}{CLIP-base} &ALL &8.13	&13.48	&7.94	&13.34  \\
    &Fig-Architecture &2.23	&3.67	&2.22	&3.86  \\
    &Fig-Illustration &4.64	&7.64	&4.66	&7.69 \\
    &Fig-Result &6.98	&11.31	&7.13	&11.74 \\
    &Table-Parameter & 1.74	&2.99	&1.68	&2.94 \\
    &Table-Result & 3.01	&5.13	&1.54	&2.85\\ \midrule
    CLIP-base+BERT &ALL & 2.47	&5.01	&3.11	&5.85\\ \midrule
    \multirow{6}{*}{BLIP-base} &ALL & 6.14	&11.30	&6.18	&11.71 \\
    &Fig-Architecture &  0.02	&0.04	&0.02	&0.02\\
    &Fig-Illustration &0.07	&0.14	&0.10	&0.17\\
    &Fig-Result & 3.26	&6.48	&3.40	&6.50\\
    &Table-Parameter  &0.01	&0.01	&0.01	&0.00  \\
    &Table-Result&0.30	&0.54	&0.30	&0.57 \\ \midrule
    BLIP-base+BERT &ALL &11.51	&20.09	&12.69	&21.77\\ \midrule
    BLIP2-FLAN-T5-XL &All &4.44 	& 7.74 	& 2.27 &4.48\\
    \bottomrule
    \end{tabular}
}
\caption{The results of fine-tuning models that are trained on different types of training data.}
\label{tab: fine-tuning model on different training data}
\end{table}

\paragraph{The effect of text-image matching task.}

As shown in Table \ref{tab: fine-tuning model on different training data}, the BLIP-2 series of models outperform other large VLMs in both Figure and Table subcategories, especially in the forward direction task. We believe that this is because BLIP-2 incorporates the text-image matching task and the image-grounded text generation task during its pre-training process to better align textual and visual information.  
The experimental results demonstrate that other models solely relying on image-grounded text generation tasks may not yield effective representations for multi-modal retrieval. Therefore, dedicated pre-training for multi-modal retrieval still requires a primary focus on the text-image matching task.

\section{Zero-shot Analysis}
\label{Appendix Zero-shot Analysis}
\paragraph{CLIP-base and BLIP-base.}
As shown in Table \ref{tab:zero-shot results of figure} and Table \ref{tab:zero-shot results of table}, the CLIP-base captioning baseline, which is specifically designed for image-text matching, shows certain generalisability in both forward and inverse retrieval across all subcategories within the Figure and Table subsets. In contrast, the BLIP-base model shows nearly no signs of effective learning on the scientific domain multi-modal data. 
These models have strong MMIR abilities for generic topic data, such as BLIP achieving an IR@1 of 86.7\% on the Flicker dataset in the zero-shot setting, whilst BLIP does not surpass 0.05\% MMR. This further demonstrates the challenges presented for MMIR in scientific domains.

\end{document}